\documentclass[twocolumn,showpacs,preprintnumbers,amsmath,amssymb]{revtex4}
\usepackage{graphicx} 
\usepackage{dcolumn}  
\usepackage{bm}       
\begin{document}
\title{Doping driven Small-to-Large Fermi surface transition and d-wave
superconductivity in a two-dimenional Kondo lattice}
\author{R. Eder$^{1}$ and P. Wr\'obel$^2$}
\author{}
\affiliation{$^1$Karlsruhe Institut of Technology,
Institut f\"ur Festk\"orperphysik, 76021 Karlsruhe, Germany\\
$^2$Institute for Low Temperature and Structure Research,
P.O. Box 1410,50-950 Wroc{\l}aw 2, Poland }
\date{\today}
\begin{abstract}
We study the two-dimensional Kondo lattice model with an additional 
Heisenberg exchange
between localized spins. In a first step we use mean-field theory with two 
order parameters. The first order parameter is a complex pairing amplitude 
between conduction electrons and localized spins which describes condensation
of Kondo (or Zhang-Rice) singlets. A nonvanishing value implies 
that the localized spins contribute to the Fermi surface volume.
The second order parameter describes singlet-pairing between the
localized spins and competes with the Kondo-pairing order parameter.
Reduction of the carrier density in the conduction band reduces the
energy gain due to the formation of the large Fermi surface and induces
a phase transition to a state with strong singlet correlations
between the localized spins and a Fermi surface which comprises only the 
conduction electrons. The model thus shows a doping-driven change of its
Fermi surface volume.
At intermediate doping and low temperature there is a phase where both 
order parameters
coexist, which has a gapped large Fermi surface and d$_{x^2-y^2}$
superconductivity. The theory thus qualitatively reproduces the phase diagram
of cuprate superconductors.
In the second part of the paper we show how the two 
phases with different  Fermi surface volume emerge in a strong coupling theory 
applicable in limit of large Kondo exchange. The large-Fermi-surface phase 
corresponds to a `vacuum' of localized Kondo singlets with uniform phase
and the quasiparticles 
are spin-1/2 charge fluctuations around this fully paired state. In the 
small-Fermi-surface phase the quasiparticles correspond to propagating 
Kondo-singlets or triplets whereby the phase of a given Kondo-singlet 
corresponds to its momentum. In this picture a phase transition occurs 
for low filling of the conduction band as well.
\end{abstract}
\pacs{71.10.Fd,74.72.-h,71.10.Ay}
\maketitle
\section{Introduction}
The existence and shape of the Fermi surface and its change with 
the hole concentration $\delta$ in the CuO$_2$ planes
appears be one of the central issues in the physics of cuprate
superconductors. In the overdoped compound Tl$_2$Ba$_2$CuO$_{6+x}$
magnetoresistance measurements\cite{Hussey},
angle-resolved photoemission
spectroscopy (ARPES)\cite{Plate} and quantum 
oscillation experiments\cite{Vignolle}
show a situation which is reminiscent of that in heavy Fermion
compounds: despite the participation of the strongly correlated
Cu3d orbitals in the states near the Fermi energy
the Fermi surface agrees well with LDA band structure
calculations which take the Cu3d-electrons as itinerant,
quantum oscillation experiments show the validity of the Fermi liquid
description with an enhanced band mass. The only moderate
mass enhancement in the cuprates thereby
seems natural given the  large Cu3d-O2p exchange constant,
$W\approx 1\;eV$, which would give a very high nominal Kondo temperature.\\
In the underdoped compounds the situation is more involved. ARPES shows
`Fermi arcs'\cite{Damascelli} which however are probably just the `visible'
part of hole pockets centered near $(\frac{\pi}{2},\frac{\pi}{2})$.
This is plausible because
the sharp drop of the ARPES weight of the  quasiparticle band
upon crossing the noninteracting Fermi surface which must be invoked
to reconcile the `Fermi arcs' with the hole pocket scenario, 
is actually well established in insulating cuprates 
such as Sr$_2$Cu$_2$O$_2$Cl$_2$\cite{Wells} and
Ca$_2$CuO$_2$Cl$_2$\cite{Ronning} where this phenomenon
has been termed the 'remnant Fermi surface'.
Meng {\em et al.} reported the observation of
the  previously unresolved `dark side' of the hole pockets
in underdoped Bi$_2$(Sr$_{2-x}$La$_x$)CuO$_6$ by ARPES\cite{Meng}.
Their conclusions subsequently were criticized\cite{King}
and the issue still seems controversial\cite{Mengreply}.\\
Moreover  both the Drude weight
in La$_{2-x}$Sr$_x$CuO$_4$\cite{Uchida,Padilla}
and YBa$_2$Cu$_3$O$_y$\cite{Padilla} as well as the inverse
low temperature Hall constant in 
La$_{2-x}$Sr$_x$CuO$_4$\cite{Ong,Takagi,Ando,Padilla} and
YBa$_2$Cu$_3$O$_y$\cite{Padilla}
scale with $\delta$
and the inferred band mass is 
constant troughout the underdoped regime and in fact even the 
antiferromagnetic phase\cite{Padilla}. 
This is exactly the behaviour expected for
hole pockets. On the other hand, 
for $\delta \ge 0.15$ the Hall constant in La$_{2-x}$Sr$_x$CuO$_4$
changes rapidly, which suggests a change from hole pockets to a large
Fermi surface\cite{Ong}.
Quantum-oscillation experiments on underdoped
YBa$_2$Cu$_3$O$_{6.5}$\cite{Doiron,Sebastian_1,Jaudet,Audouard}
and YBa$_2$Cu$_4$O$_8$\cite{Yelland,Bangura}
show that the Fermi surface has a cross section that is comparable 
to $\delta/2$ rather than $(1-\delta)/2$ as in the ovderdoped
compounds. Here it should be noted that the mere
validity of the Fermi liquid description as demonstrated by the 
quantum oscillations is conclusive evidence against the
notion of `Fermi arcs': the defining property of a Fermi liquid is the
one-to-one correspondence of its low-lying states to those of
a fictitious system of weakly interacting Fermionic quasiparticles
and the Fermi surface of these quasiparticles
is a constant energy contour of their dispersion
and therefore necessarily a closed curve in ${\bf k}$-space.
On the other hand the quantum oscillations cannot be viewed
as evidence for hole pockets either in that
both the Hall constant\cite{LeBoeuf} and thermopower\cite{Chang} have a sign 
that would indicate electron pockets. Thereby
both, the Hall constant and the thermopower, show a strong
temperature dependence and in fact a sign change
as a function of temperature. This sign change is observed only at
where superconductivity is suppressed by
a high magnetic field. At the same
time neutron scattering experiments on
YBa$_2$Cu$_3$O$_{6.6}$ in the superconducting state
show pronounced anisotropy in the spin
excitations spectrum below $30\;meV$ and at low temperatures\cite{Hinkov}.
This indicates an as yet not fully understood anisotropic state,
possibly to a `nematic' state with inequivalent
$x$- and $y$-direction in the CuO$_2$ plane. Such a
nematicity has also been observed in scanning tunneling 
microscopy experiments on Bi$_2$Sr$_2$CaCu$_2$O$_{8+\delta}$\cite{Lawler}
and will modify the Fermi surface in some way which may explain the
unexpected sign.
More recently, Sebastian {\em et al.} concluded from an analysis of the 
2$^{nd}$ harmonic in quantum oscillations in underdoped
YBa$_2$Cu$_3$O$_{6+x}$ that the Fermi surface consists only of a single 
pocket\cite{Sebationsinglepock}. Since this would rule out the possibility 
of coexisting hole-like and electron-like Fermi surface 
sheets\cite{Chakravarty} and since it is hard to imagine that the sole 
Fermi surface sheet of a hole-doped compound is electron-like
this result would imply that the Fermi surface actually is
a hole pocket and that the sign of the Hall constant and thermopower
does not reflect the nature of the carriers but is determined by
some other mechanism. Adopting this point of view,
the picture of hole pockets centered near 
$(\frac{\pi}{2},\frac{\pi}{2})$ with an area $\propto \delta$
would give a simple
and consistent description of ARPES, normal state Drude weight and
Hall constant and quantum oscillation experiments in the
underdoped state. Combined with the results for the overdoped
compounds this would imply that as a function of
 $\delta$ the cuprates undergo
a phase transition between two states with different
Fermi surface  volume, whereby the Cu 3d electrons
contribute to the Fermi surface volume in the overdoped compounds but
`drop out' of the Fermi surface volume in the underdoped regime.\\
Interestingly, the superconducting transition in the cuprates
itself seems to be accompanied by a Fermi surface
change as well. Namely ARPES shows that the quasiparticle
peaks near $(\pi,0)$, which are hardly distinguishable in the normal
state, become very intense and sharp in the superconducting 
state\cite{Dessau,Hwu}. This looks as if coherent
quasiparticles around $(\pi,0)$ exist only in
the superconducting state and in fact seem to jump into existence
right at the superconducting transition\cite{ShenSawatzky}.
A possible interpretation would be that the superconducting
transition occurs between a hole pocket-like
Fermi surface - which does not extend towards $(\pi,0)$ - to a gapped
large Fermi surface which naturally has some portions near  $(\pi,0)$.\\
Transitions where the correlated electron subsystem contributes to the
Fermi surface volume or not are not unfamiliar in Heavy
Fermion compounds. An example is the metamagnetic transition in
CeRu$_2$Si$_2$
where the Ce 4f-electrons, which contribute to the Fermi surface
in zero magnetic field, seem to drop out of the Fermi
volume as the magnetic field is increased\cite{Aoki}.
In CeRh$_{1-x}$Co$_x$In$_5$ the Ce 4f electrons change from
localized for $x\le 0.40$ to itinerant for $x\ge 0.50$
as the lattice constant decreases due to substitution of
Rh by the smaller Co\cite{Goh}. 
Similarly, the localized Ce 4f electrons in CeRh$_2$Si$_2$\cite{Onuki},
CeRhIn$_5$\cite{Shishido} and CeIn$_3$\cite{Settai}
can be made itinerant by pressure.
It seems that in the case of the metamagnetic transition
the magnetic field
breaks the Kondo singlets between Ce 4f spins and conduction
electrons whereas in the other cases the decrease of
the hybridization strength between 4f and conduction electrons 
makes the formation of Kondo singlets unfavourable.\\
It has been pointed out long ago by Doniach
that there may be a competition between the Kondo effect
and the RKKY interaction which then leads to a phase transition to a
magnetically ordered phase if certain parameters in the system are
varied\cite{Doniach}. More recently Senthil {\em et al.} have investigated this
question in the context of Heavy Fermion compounds and 
discussed a transition to a magnetically ordered state which is
accompanied by a Fermi surface transition from large to small
as the strength of the Kondo coupling is varied\cite{Senthil}.\\
In the cuprates one might expect another reason for a
Fermi surface transition, namely the
depletion of the mobile carriers i.e. holes
in O2p orbitals. It is self-evident that the gain in energy
due to formation of a common Fermi sea of mobile
O2p holes and localized Cu3d spins
must tend to zero when the density of mobile carriers vanishes.
More precisely, one might expect the band filling to play a substantial
role when the width of the occupied part of the conduction
band becomes smaller than the width of the
Kondo resonance at the Fermi level. Since the
Cu3d-O2p exchange constant is large and the band filling small
this situation may well be realized in the cuprates. As the
density of holes in O2p orbitals is reduced one would thus
expect that at some point the
localized spins drop out of the Fermi surface so as to optimize
their mutual superexchange energy instead. Due to the 
near-two-dimensionality of the cuprates, however, the transition
is not to a magnetically ordered state, but to a `spin liquid'
with strong nearest-neighbor singlet correlations instead of
true antiferromagnetic order.\\
Cuprate superconductors are frequently described by a single band
Hubbard model or the t-J model. Whereas a Kondo-lattice-like Hamiltonian
can be derived by lowest order canonical perturbation model
from the so-called d-p model for the CuO$_2$ plane\cite{ZaanenOles}
these single band models are obtained in a subsequent step which is
valid in the limit of large Kondo coupling between O2p and Cu3d
electrons, so that the Kondo singlet (Zhang-Rice-singlet)
extends over little more than only one plaquette. Yet, these
models should show a Fermi surface transition as well if they
are equivalent to the CuO$_2$ plane.
Exact diagonalization studies of the t-J model
have indeed shown that the Fermi surface at hole dopings
$\le 15\%$ takes the form of hole pockets\cite{poc1,poc2,poc3},
that the quasiparticles have the character of strongly renormalized
spin polarons throughout this doping range\cite{r1,r2,r3} and that the low 
energy spectrum at these doping levels can be described as a Fermi liquid of 
spin $1/2$ quasiparticles corresponding to the doped holes\cite{lan}.
A comparison of the dynamical spin and density correlation function
at low\cite{den1,den} ($\delta < 15\%$)
and intermediate  and high ($\delta=30-50 \%$) hole doping moreover
indicates\cite{intermediate} that around optimal doping a phase 
transition to a state with large Fermi surface takes place in the
t-J model. A study of the electronic self-energy in the
single-band Hubbard model has indicated that there such a transition
takes place as well\cite{seki}. Contrary to widespread belief
such hole pockets can be completely consistent with the
Luttinger theorem for the single band Hubbard model\cite{seki}.\\
To study the  issue of the Fermi surface transition
further we performed a Hartree-Fock treatment of a
two-dimensional Kondo lattice with an additional Heisenberg
exchaneg between localized spins. This is presented in Section II.
Interestingly it turns out that such a transition not only exists
but is generically accompanied by $d_{x^2-y^2}$ superconductivity
in the mobile carrier system
which might provide an explanation for the phase
diagram of the cuprates. In simplest terms
superconductivity occurs because the coherent 
Kondo-pairing between $d$-spins and $c$-electrons `transfers' the singlet
pairing between the localized $d$-spins to the mobile $c$-electron sytem.\\
Since a mean-field treatment may not really be expected to be valid
in the limit of large superexchange $W$ between localized and conduction 
electrons, which is the 
region of physical interest for the cuprates, we also 
show how the two phases with small and large
Fermi surface may be understood in the limit of large $W$.
The small Fermi surface phase is discussed in section III
and is similar to the lightly doped Mott insulator
as discussed in Ref. \cite{pockets}. 
The phase with large Fermi surface
is discussed in section IV, closely following Refs. \cite{kondo_1}
and \cite{kondo_2}. 
Section V discusses the possibility of a phase transition
between small and large fermi surface within the strong coupling
theory, Section VI gives a summary and discussion.
\section{Mean field Theory}
\begin{figure}
\includegraphics[width=\columnwidth]{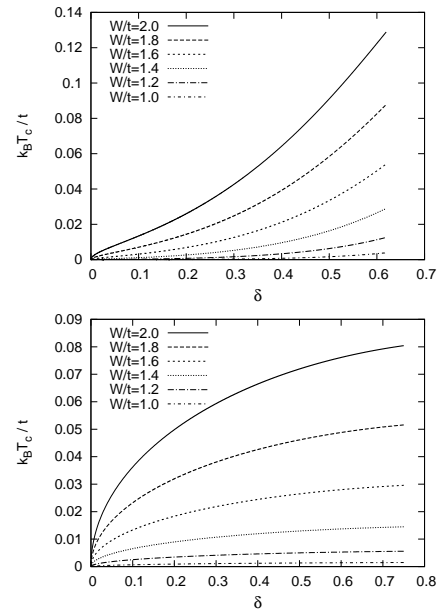}
\caption{\label{fig14}  
Critical temperature for the onset of the parameter
$\Delta_{cd}$ with the true 2-dimensional nearest-neighbor-hopping
band structure (top) and a constant density of states
in the Intervall $[-4t:4t]$ (bottom).
}
\end{figure}

\begin{figure}
\includegraphics[width=\columnwidth]{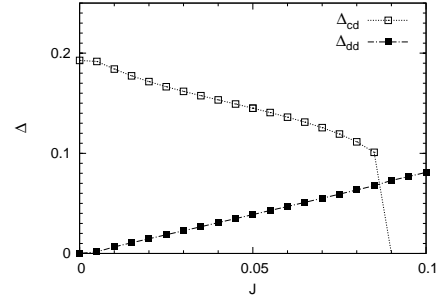}
\caption{\label{fig13}  
Development of the two order parameters with increasing $J$.
The other values are $W/t=1.6$, $\delta=0.2$ and $T/t=0.001$.
}
\end{figure}

We study a Kondo lattice model which consists of
a single metallic conduction band - described by the Fermionic operators
$c_{i,\sigma}^\dagger$ - and a lattice of localized spins, described
by $d_{i,\sigma}^\dagger$. The lattice sites - labelled by $i$ -
form a simple cubic lattice and
there is one localized spin and one $s$-like conduction
orbital in each unit cell.
A similar model with two $O2p$-like conduction orbitals per unit cell
can be derived by canonical transformation for the CuO$_2$ 
plane\cite{ZaanenOles}.
We augment the model by a direct Heisenberg exchange between nearest neighbor
localized spins. A similar model for the CuO$_2$ plane
has been studied previously\cite{Fedro}.
To make more easy contact with the CuO$_2$ planes
we consider the operators $c_{i,\sigma}^\dagger$ and $d_{i,\sigma}^\dagger$
to create holes rather than electrons.
We denote the density of holes in the conduction band
by $\delta$ - the total densisty of holes/unit cell thus is
$1+\delta$. The Hamiltonian reads
\begin{eqnarray}
H&=&H_t + H_W + H_J,\nonumber \\
H_t &=& t \sum_{\langle i,j\rangle}
\sum_\sigma\;c_{i,\sigma}^\dagger c_{j,\sigma}^{},
\nonumber \\
H_W &=& W \sum_i \left(\vec{S}_{d,i}\cdot \vec{S}_{c,i} 
- \frac{n_{c,i} n_{d,i}}{4}\right), \nonumber \\
H_J &=& J \sum_{\langle i,j\rangle } \left( \vec{S}_{d,i}\cdot \vec{S}_{d,j}
- \frac{n_{d,i} n_{d,j}}{4}\right), 
\label{hamil}
\end{eqnarray} 
where 
\begin{eqnarray}
 \vec{S}_{c,i}&=&\frac{1}{2} \;c_{i,\alpha}^\dagger \vec{\sigma}_{\alpha,\beta}^{} 
c_{i,\beta}^{}, \nonumber \\
n_{c,i}&=&c_{i,\alpha}^\dagger c_{i,\alpha}^{},
\end{eqnarray}
with $\vec{\sigma}$ the vector of the Pauli matrices
and analogous definitions hold for $\vec{S}_{d,i}$ and $n_{d,i}$.
The model is to be considered in the sector of the Hilbert space
where all $d$-orbitals are singly occupied.\\
In a first step we drop the $d-d$ exchange $H_J$.
We use the identity
\begin{eqnarray}
\vec{S}_{d,j}\cdot \vec{S}_{c,j}- \frac{n_{d,j} n_{c,j}}{4} &=&
-\frac{1}{2} \;S_j^\dagger\; S_j^{},
\nonumber \\
S_j^\dagger &=& c_{j,\uparrow}^\dagger d_{j,\downarrow}^\dagger -
c_{j,\downarrow}^\dagger d_{j,\uparrow}^\dagger\;,
\end{eqnarray}
and apply the Hartee-Fock approximation
\begin{equation}
S_j^\dagger\; S_j^{} \approx \langle S_j^\dagger \rangle S_j^{}
+ S_j^\dagger\; \langle S_j^{} \rangle - \langle S_j^\dagger \rangle
 \langle S_j^{} \rangle.
\end{equation}
Whereas the orginal Hamiltonian conserves the number of
d-holes at each site, this does not hold true for the
mean-field Hamiltonian - which is clearly a severe drawback of the
theory. Although the present decoupling scheme is different from
that used by Senthil {\em et al.}\cite{Senthil} the theories can be converted
into each in the case $J=0$ by performing
a particle-hole transformation for the $d$-electrons.\\
There are two side
conditions to be obeyed: one for the total hole number, $N_h$,
the other one for the number of $d$-holes, $N_d$, which must be equal to
$N$, the number of $d$-sites in the system. We enforce these by Lagrange
multipliers $\mu$ and $\lambda$, respectively, and,
introducing the vector
$v_{{\bf k},\sigma}^\dagger=(c_{{\bf k},\sigma}^\dagger,
d_{{\bf k},\bar{\sigma}})$, we
obtain the Fourier transformed Hamiltonian
\begin{eqnarray}
H - \mu N_h - \lambda N_d &=& \sum_{{\bf k},\sigma}\;v_{{\bf k},\sigma}^\dagger\;H_{{\bf k},\sigma}^{}
v_{{\bf k},\sigma}^{}  \nonumber \\
&&\;\;- 2N \epsilon_d + 2N\mu + N\frac{\Delta_{cd}^2}{2W},\nonumber\\
\end{eqnarray}
where $N_h$ ($N_d$) are the operators for the total number of holes
(number of $d$-like holes), $N$ is the number of sites, and
\begin{eqnarray}
H_{{\bf k},\sigma}&=&
\left(\begin{array}{c c}
\epsilon_{\bf k} - \mu & -sign(\sigma) \frac{\Delta_{cd}}{2}, \\
-sign(\sigma)\frac{\Delta_{cd}^*}{2} & \epsilon_d - \mu
\end{array} \right),
\nonumber \\
\Delta_{cd} &=& W \langle S_j \rangle \\
\epsilon_d &=& \lambda + 2 \mu.
\end{eqnarray}
The parameter $\Delta_{cd}$ describes coherent singlet formation
between the conduction electrons and the localized spins.
The essence of the Kondo-effect is the quenching of
the magnetic moments that means a localized spin at site $i$ forms
a singlet with a conduction electron (this would be the Zhang-Rice
singlet in the case of cuprate superconductors). The phase of
this local singlet then is in principle undetermined
and the above mean-field decoupling describes a state, where this
phase is uniform over the whole system.\\
Next,  being the expectation value
of two creation/annihilation operators, $\Delta_{cd}$
has some similarity to a superconducting order
parameter. Since the pairing is not between time-reversed states,
however, the resulting gound state is not superconducting.
This will be apparent from the fact that it has a well-defined
Fermi surface.\\
The Hamiltonian can be diagonalized by the transformation
\begin{eqnarray}
\gamma_{{\bf k},1,\sigma}^\dagger &=&\;
u_{\bf k} c_{{\bf k},\sigma}^\dagger + sign(\sigma)
v_{\bf k} d_{-{\bf k},\bar{\sigma}}^{},
\nonumber \\
\gamma_{{\bf k},2,\sigma}^\dagger &=&
-sign(\sigma)v_{\bf k} c_{{\bf k},\sigma}^\dagger + u_{\bf k} d_{-{\bf k},\bar{\sigma}}^{},
\end{eqnarray}
and we obtain the two quasiparticle bands
\begin{eqnarray}
E_{\pm,{\bf k}}&=&\frac{1}{2}
\left( \epsilon_{\bf k} + \epsilon_d \pm W_{\bf k}\right)-\mu,
\nonumber \\
W_{\bf k} &=& \sqrt{(\epsilon_{\bf k} - \epsilon_d)^2 + \Delta_{cd}^2},
\nonumber \\
u_{\bf k} &=& \left(\frac{1}{2} - \frac{\epsilon_{\bf k} - \epsilon_d}{2W_{\bf k}}\right)^{1/2},
\nonumber \\
v_{\bf k} &=& \left(\frac{1}{2} + \frac{\epsilon_{\bf k} - 
\epsilon_d}{2W_{\bf k}}\right)^{1/2}.
\end{eqnarray}
In the equation for $v_{\bf k}$ it is assumed that $\Delta_{cd} > 0$.
The self-consistency equation for $\Delta_{cd}$ becomes
\begin{equation}
1=\frac{W}{N}\sum_{\bf k}
\frac{1}{W_{\bf k}}\frac{\sinh\left(\frac{\beta W_{\bf k}}{2}\right)}
{ \cosh\left(\frac{\beta W_{\bf k}}{2}\right) + 
\cosh\left(\beta \left(\frac{\epsilon_{\bf k} +
  \epsilon_d}{2}-\mu\right)\right)}.
\label{selfco}
\end{equation}
Next, it is straightforward to show that
\begin{equation}
2N - N_d + N_c  = \sum_{{\bf k},\sigma}\sum_{\nu=1}^2
\gamma_{{\bf k},\nu,\sigma}^\dagger\gamma_{{\bf k},\nu,\sigma}^{}
\end{equation}
where $N_d$ ($N_c$) are the operators for the number of
$c$-like ($d$-like) holes. If we demand that $N_d=N$
the operator $N_h$ of the total hole number becomes
\begin{equation}
N_h = \sum_{{\bf k},\sigma}\sum_{\nu=1}^3
\gamma_{{\bf k},\nu,\sigma}^\dagger\gamma_{{\bf k},\nu,\sigma}^{}
\end{equation}
so that we have a `large' Fermi surface which comprises the
localized $d$-holes. This, however, will hold only
if there is exactly one localized spin/d-site.
When $\Delta_{cd}$ is zero the $c$-holes and $d$-holes are decoupled
and the condition  $N_d=N$ puts a dispersionless $d$-band right
at $\mu$, i.e. $\epsilon_d=\mu$. The Fermi surface then is that
of the decoupled $c$-holes and accordingly has
a volume which does not comprise the $d$-holes.\\
As already stated for
$\Delta_{cd}\rightarrow 0$ we have $\epsilon_d \rightarrow \mu$
and inserting this into (\ref{selfco})
we obtain the equation for the critical temperature
\begin{equation}
1=\frac{W}{2N}\sum_{\bf k}\frac{1}{\epsilon_{\bf k}-\mu}
\tanh\left(\frac{\beta_c (\epsilon_{\bf k}-\mu)}{2}\right).
\label{tceq}
\end{equation}
Figure \ref{fig14} shows $T_c$ as a function of $\delta$.
This is shown both for the true 2-dimensional nearest-neighbor-hopping
dispersion and for a conduction band with a constant density of states
in the interval $[-4t:4t]$. It is obvious that $T_c \rightarrow 0$
as $\delta\rightarrow 0$. In the limit of small $T_c$ we can
obtain a rough approximation by replacing the integrand in
(\ref{tceq}) by $(|\epsilon_{\bf k}-\mu|+T_c/2)^{-1}$.
For the constant density of states we find in this way
\begin{equation}
T_c= 2t\;e^{-8t/W}\sqrt{\delta(2-\delta)}.
\end{equation}
As expected, the energy gain due to
the formation of common large Fermi surface thus goes to zero
when the density of mobile carriers vanishes. For sufficiently
low temperature the large Fermi surface thus is
formed for any carrier concentration, but if there is a
competing term in the Hamiltonian there may be a phase transition
at finite doping.\\
We now introduce the $d-d$ Heisenberg exchange $\propto J$.
We decouple the $d$-$d$ Heisenberg exchange in the same way as the
$c$-$d$ exchange:
\begin{eqnarray*}
\vec{S}_{d,i}\cdot \vec{S}_{d,j}- \frac{n_{d,i} n_{d,j}}{4}
&=& -\frac{1}{2} s_{ij}^\dagger s_{ij}\nonumber \\
s_{ij}^\dagger &= &
d_{i,\uparrow}^\dagger d_{j,\downarrow}^\dagger -
d_{i,\downarrow}^\dagger d_{j,\uparrow}^\dagger \nonumber \\
s_{ij}^\dagger s_{ij} &\approx&
\langle s_{ij}^\dagger \rangle  s_{ij} +  s_{ij}^\dagger \langle  s_{ij}\rangle
- \langle s_{ij}^\dagger \rangle \langle  s_{ij}\rangle
\end{eqnarray*}
Inclusion of this term
doubles the dimension of the matrices to be considered.
Introducing
$v_{\bf k}=(c_{{\bf k},\uparrow}^\dagger, d_{-{\bf k},\downarrow}^{}, 
c_{-{\bf k},\downarrow}^{}, d_{{\bf k},\uparrow}^\dagger)$
the Hamilton matrix becomes
\begin{eqnarray}
H&=& \sum_{{\bf k}}\;v_{{\bf k}}^\dagger\;H_{{\bf k}}v_{{\bf k}}
\nonumber \\
H_{{\bf k}}&=& \left( \begin{array}{ c c c c }
\epsilon_{\bf k}-\mu & -\frac{\Delta_{cd}}{2} & 0 & \\
-\frac{\Delta_{cd}^*}{2} & \epsilon_d-\mu & 0 & -\Delta_{dd}\gamma_{\bf k} \\
0 & 0 & -\epsilon_{\bf k}+\mu & -\frac{\Delta_{cd}}{2} \\
0 & -\Delta_{dd}\gamma_{\bf k} & -\frac{\Delta_{cd}^*}{2} & -\epsilon_d+\mu 
\end{array}\right)
\nonumber \\
\gamma_{\bf k} &=& \frac{1}{2}\left(\cos(k_x)\pm \cos(k_y)\right)
\nonumber \\
\Delta_{dd}&=& \frac{zJ}{2}\langle s_{ij}\rangle
\label{bigmfham}
\end{eqnarray}
The two parameters, $\Delta_{cd}$ and $\Delta_{dd}$ now have to
be determined self-consistently. Unlike the on-site order parameter
$\Delta_{cd}$, the $d-d$ pairing amplitude is a `bond related'
quantity and thus may have different sign
for bonds along $x$ and $y$ so that we may have
$s$-like pairing or $d$-like pairing. These two possibilities
have to be considered separately.\\
Before studying the full problem we briefly consider the case
$\Delta_{cd}=0$. In this case the $d$- and $c$-holes are again
decoupled and we only have to treat the $d$-electron system.
The chemical potential $\mu$ is determined by the conduction
holes alone and we must have $\epsilon_d=\mu$.
The result will not depend on $\delta$
or $W$ and the temperature dependence is universal if
temperature is measured in units of $J$. Finally, there is
no difference between $\gamma_{s,d}({\bf k})$ because
$\cos(k_y) = - \cos(k_y+\pi)$ so that the $d$-like pairing
simply corresponds to a shift of the Brillouin
zone by $(\pi,0)$. 
The self-consistency equation reads
\begin{equation}
\Delta_{dd} = \frac{2J}{N}\sum_{\bf k}\;
\gamma({\bf k}) \;\frac{\sinh(\beta \gamma({\bf k}) \Delta_{dd})}
{1 + \cosh(\beta \gamma({\bf k}) \Delta_{dd})}
\end{equation}
and the temperature for the phase transition is $\tilde{T}_c=J/4$.
The gap at $T=0$ is found to be
\begin{eqnarray}
\Delta_{dd} &=& J \sum_{\bf k} |\gamma({\bf k})|
\nonumber \\
&=& J\cdot 0.81.
\end{eqnarray}
The localized spins aquire a nonvanishing dispersion, which is
unphysical and an artefact of the mean-field 
approximation. The problem is somewhat lessened in that the
localized electrons at least have no Fermi surface.
To see this we note first that the dispersion consists of two
bands $\epsilon_d \pm \Delta_{dd}\gamma({\bf k})$. Since $\epsilon_d=\mu$
the lower of these is completely filled.
The momentum distribution function for the $d$-electrons then becomes
\begin{eqnarray}
n_{d,\bf{k}} &=& \sum_{\sigma}
\langle d_{{\bf k},\sigma}^\dagger d_{{\bf k},\sigma}^{} \rangle
\nonumber \\
&=& 1
\end{eqnarray}
so that at least the momentum distribution of the $d$-electrons
is consistent with localized electrons.\\
Next we switch to the full problem of two coupled order parameters
i.e. $\Delta_{cd}\ne 0$ and $\Delta_{dd}\ne 0$.
All results presented below have been obtained by numerical solution
of the self-consistency equations on a $400\time 400$ lattice
with periodic boundary conditions. Study of the variation with lattice size
shows that this implies a reasonable convergence.\\
As already mentioned $\Delta_{dd}$ may be 
s-like and d$_{x^2-y^2}$-like. This difference
will now matter because the relative position of the lines of zeroes 
in the $d$-electron dispersion and the Fermi surface
of the $c$-electrons makes a physical difference.
To decide which of the two symmetries is realized we consider
the free energy/site given by
\begin{eqnarray}
f(T,n)&=& -\frac{1}{\beta}
\sum_{{\bf k}}\sum_{\nu=1}^4 \;\log\left(1+ e^{-\beta E_{{\bf k},\nu}}\right)
\nonumber \\
&& \;\;\;\;\;\;\;\;+\frac{\Delta_{cd}^2}{2W} +
\frac{\Delta_{dd}^2}{4J} +(\delta-1) \mu.
\end{eqnarray}
Numerical evaluation shows
that the s-like state is never realized.\\
Figure \ref{fig13} then shows an example of the development of the
two coupled order parameters as $J$ is switched on.
It is quite obvious that increasing $\Delta_{dd}$ is detrimental
to $\Delta_{cd}$ i. e. the two order parameters
are competing. Nonvanishing $J$ thus may introduce
a phase transition at $T=0$ as a function of doping between the two phases.\\
To discuss the phase diagram we note first that we can distinguish
two regimes: at low doping the critical temperature $T_c$
for the onset of
$\Delta_{cd}$ will be below $J/4$, which is the critical temperature
for the onset of $\Delta_{dd}$.
This means that as the temperature is
lowered, a nonvanishing $\Delta_{dd}$ sets in first and
$\Delta_{cd}$ then must develop on the `background' of this nonvanishing
$\Delta_{dd}$. When $T_c > J/4$ on the other hand, we first have a nonvanishing
$\Delta_{cd}$ and $\Delta_{dd}$ develops at lower temperature.
As will be seen next, the two regimes are more different than might be
expected at first sight. For the time being
we fix $W/t=1.6$ and $J/t=0.05$. As can be seen 
in Figure \ref{fig14},
the doping where $T_c=J/4$ then is approximately $\delta=0.3$.
\begin{figure}
\includegraphics[width=\columnwidth]{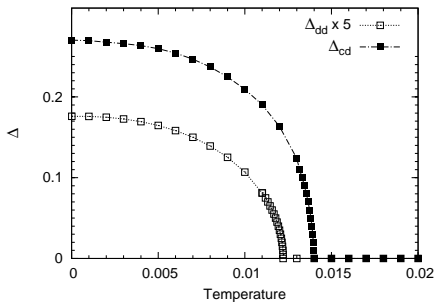}
\caption{\label{fig8}  
Temperature dependence of the two order parameters for
$W/t=1.6$, $J/t=0.05$, $\delta=0.32$.}
\end{figure}
Figure \ref{fig8} then shows the self-consistent $\Delta$'s 
as a function of 
temperature for $\delta=0.32$. Both parameters show a fairly conventional
behaviour at the two phase transitions with the
characteristic $\sqrt{T_c-T}$ behaviour
and the same is seen whenever $\Delta_{cd}$ sets in 
at higher temperature.\\
The situation is quite different for $\delta<0.3$ 
as can be seen in Figure \ref{fig6}. For most dopings
there is now a small but finite temperature
range where two solutions $(\Delta_{cd}\ne 0,\Delta_{dd}\ne 0)$ exist
(in addition there are the solutions with $\Delta_{dd}=0$ and
$\Delta_{cd}=0$). Calculation of the free energy shows, that it is
always the solution with the higher $\Delta_{cd}$ which has the
lower free energy, i.e. this is the physical state.
\begin{figure}
\includegraphics[width=\columnwidth]{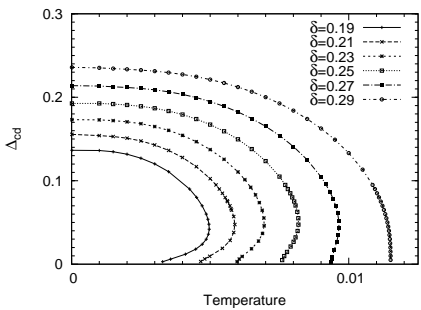}
\caption{\label{fig6}  
Self-consistent solution for $\Delta_{cd}$ as a function
of temperature for various dopings $\delta$.
The other values are $W/t=1.6$, $J/t=0.05$.
}
\end{figure}
\begin{figure}
\includegraphics[width=\columnwidth]{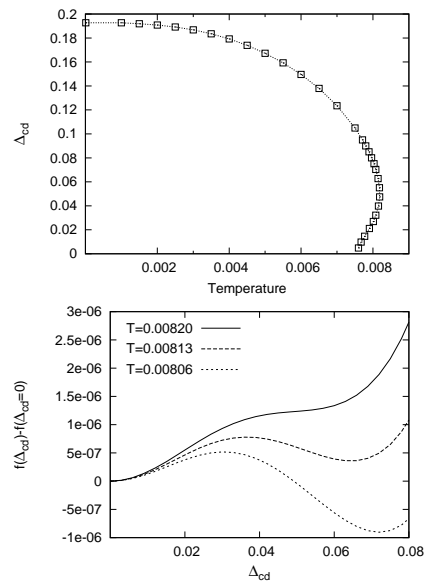}
\caption{\label{fig7}  
Top: Self-consistent solution for $\Delta_{cd}$ as a function
of temperature for $\delta=0.25$, see Figure \ref{fig6}.
Bottom: Scans of the free energy as a function of $\Delta_{cd}$
whereby all other mean-field parameters have been obtained self-consistently.
}
\end{figure}
To clarify the nature of the phase transition we fix
$\Delta_{cd}$, determine all other parameters 
$(\mu,\epsilon_d,\Delta_{dd})$ self-consistently
and evaluate the free energy $f$ as a function of
$\Delta_{cd}$. 
Points on the resulting curve $f(\Delta_{cd})$
which are stationary with respect
to variations of $\Delta_{cd}$ are stationary
with respect to variations of all parameters and therefore are
solutions to the self-consistency equations.
The result is shown in Figure  \ref{fig7}.
At high temperature $f(\Delta_{cd})$ has only one extremum,
namely a minimum at $\Delta_{cd}=0$. As the temperature is lowered,
however, one can recognize a `wiggle' in the curve which
develops into a maximum and a minimum. These
correspond to the two solutions
with nonvanishing $\Delta_{cd}$. Next there occurs a level
crossing between the two minima - we thus have a first order transition -
and as the temperature is lowered further the maximum
for $\Delta_{cd}\ne0$ `absorbs' the minimum at  $\Delta_{cd}=0$.
From then on we have only
the minimum with $\Delta_{cd}\ne 0$ and the maximum at $\Delta_{cd}=0$.
This behaviour can be seen at almost all dopings below $\tilde{\delta}$.
It is only very close to $\tilde{\delta}$ that there is only one 
solution but the 
$\Delta_{cd}(T)$ curve is very steep as $\Delta_{cd} \rightarrow 0$.
The important finding then is that the temperature-induced
transition between small and large Fermi surface
is a 1$^{st}$ order transition in this doping range.
This is not surprising in that a  2$^{nd}$ order transition 
involving competing order parameters may become 1$^{st}$ order\cite{She}.\\
\begin{figure}
\includegraphics[width=\columnwidth]{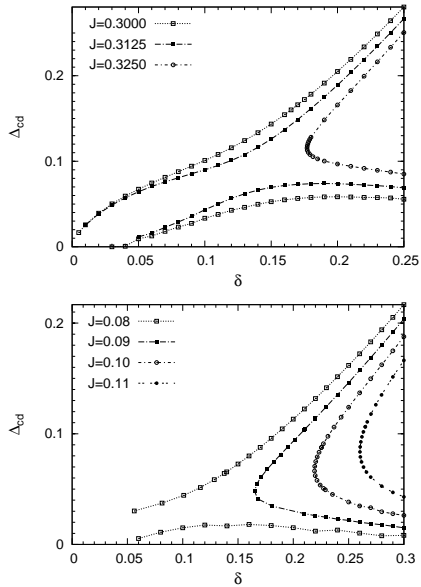}
\caption{\label{fig9}  
Self-consistent solutions for $\Delta_{cd}$ as a function
of doping for $T=0$ and different $J$. The value $W/t=2$ ($W/t=1.6$)
in the upper (lower) part of the figure.}
\end{figure}
Next we consider the dependence on hole doping $\delta$,
in particular the question if the Fermi surface transition
now occurs at finite doping even at zero temperature.
Figure \ref{fig9}, which shows the 
self consistent solutions for $\Delta_{cd}$ at $T=0$ for different
$W/t$ and $J/t$ demonstrates
that the answer depends on the magnitude of $J/t$.
For small $J$ a nonvanishing solution exists down to $\delta=0$.
In fact there are two such solutions, but computation
of the free energy shows that only the solution with the larger
$\Delta_{cd}$ is a minimum and moreover has a lower free energy than the
solution with $\Delta_{cd}=0$. Accordingly, there is no phase transition
down to $\delta=0$ for small $J/t$. If $J$ increases, 
however, there is now an extended range of small $\delta$
where no solution  with $\Delta_{cd}\ne 0$ exists. 
As can be seen in the Figure \ref{fig9}  this change occurs 
in that the maximum and minimum of $f(\Delta_{cd})$ `merge' for some
$\delta$. The hourglass shape formed by the two $\Delta_{cd}(\delta)$
curves immediately below the critical $J/t$ can be clearly seen in the
upper part of the Figure. It has been verified, however, that for 
large values of
$J/t$ where the transition has occured
there is no more solution for low
$\delta$. This means that
for larger $J/t$ a phase transition does occur and there is
a considerable range of $\delta$ where the mean-field theory predicts
a spin-liquid (i.e. $\Delta_{dd}\ne 0$)
with a small Fermi surface.\\
We proceed to a discussion of the phase diagram whereby we consider the
more interesting case of larger $J/t$. An example is
shown in Figure \ref{fig5}, the band structures for the various
phases are shown in Figures \ref{fig1} and \ref{fig2}.
\begin{figure}
\includegraphics[width=\columnwidth]{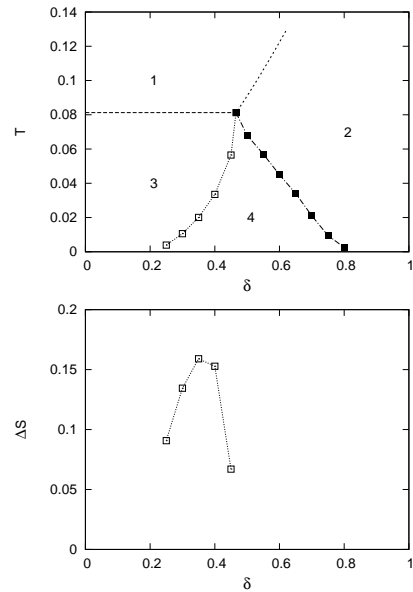}
\caption{\label{fig5}
Top: Mean-field phase diagram for the parameter values
$W/t=2$, $J/t=0.325$.\\
Bottom: Difference in entropy/site between the phases 3 and 4.
}
\end{figure}
To begin with, we can distinguish four phases. 
At high temperature (phase 1) both self-consistent parameters are zero which
implies that the conduction holes are decoupled from
the localized spins and the localized spins themselves are
uncorrelated. At higher doping and low temperature
there is a phase with
$\Delta_{cd}\ne 0$, $\Delta_{cd}= 0$ (phase 2). The band structure
(see Figure \ref{fig1}, Top part) shows that
there is a large Fermi surface
with an enhanced band mass.
This phase most likely corresponds
to the overdoped regime in cuprate superconductors.\\
For low doping and low temperature, on the other hand,
there is a phase with $\Delta_{dd}\ne 0$, $\Delta_{cd}= 0$ 
(phase 3). This has strong singlet
correlations between the localized spins
and a small Fermi surface which is centered at
$(\pi,\pi)$ in mean-field theory.
(see Figure \ref{fig1}, Bottom part). This phase likely corresponds to 
the `pseudogap' phase in 
cuprate superconductors. Thereby the strongest deficiency of the 
mean-field theory consists in the neglect of any correlation between the
localized spins and the conduction holes. One may expect that
coupling of the conduction hole pocket around $(\pi,\pi)$
to the antiferromagnetic spin
fluctuations of the localized spins will create 
hole pockets centered near $(\frac{\pi}{2},\frac{\pi}{2})$.
In any way, however, the Fermi surface does not comprise the
$d$-electrons in this phase.\\
Finally, at intermediate doping and low temperature
there is a `dome'-like region in the phase diagram 
in which both self-consistent parameters
are different from zero (phase 4). Inspection of the band
structure in Figure \ref{fig2} shows  
\begin{figure}
\includegraphics[width=\columnwidth]{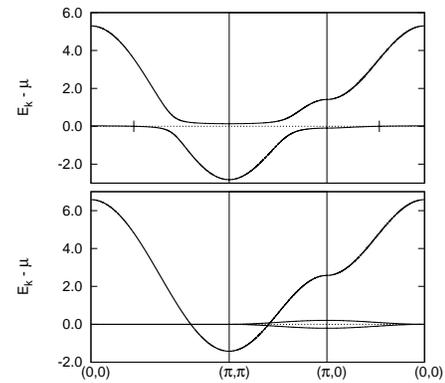}
\caption{\label{fig1}  
Top: Band structure for phase 2 ($T=0.005$, $\delta=0.60$).
The two vertical lines at $\mu$ give the Fermi momenta, i.e.
the crossings of the quasiparticle band through $\mu$.\\
Bottom: Band structure for phase 3 ($T=0.005$, $\Delta=0.25$).
Other parameter values are $W/t=2$, $J/t=0.325$.}
\end{figure}
\begin{figure}
\includegraphics[width=\columnwidth]{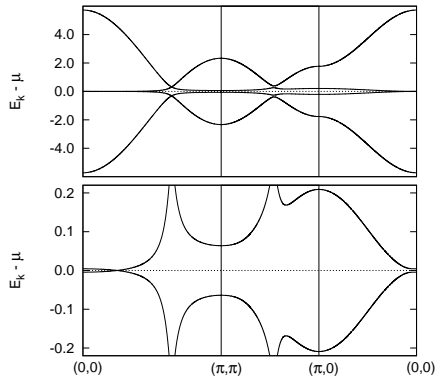}
\caption{\label{fig2}  
Band structure for phase 4 ($T=0.005$, $\delta=0.45$).
Entire bandwidth (top) and closeup of the region near $\mu$ (bottom).
Other parameter values are $W/t=2$, $J/t=0.325$.}
\end{figure}
that phase 4 does not have a Fermi surface, but rather
a gap with a node along the $(1,1)$ direction, i.e. the
band structure expected for a superconductor with
d$_{x^2-y^2}$ order parameter. The Fermi momentum along the
$(1,1)$ direction coincides with that of the large Fermi surface,
however, so that we have a gapped large Fermi surface in this
phase. Across the phase transition
on the low doping part of the dome the Fermi surface thus changes from
a hole pocket centered on $(\pi,\pi)$ to a gapped
large Fermi surface, whereas the transition on the high doping side
of the dome corresponds to the opening of a d$_{x^2-y^2}$-like
gap on the large Fermi surface. Accordingly, we consider the question
whether phase 4 is superconducting. To that end we
consider the $c$-like pairing amplitude
\begin{equation}
\Delta_{cc}=\frac{1}{N}
\sum_{\bf k}\gamma({\bf k}) \langle
c_{{\bf k}\uparrow}^\dagger c_{-{\bf k}\downarrow}^\dagger 
+ c_{-{\bf k}\downarrow}^{}c_{{\bf k}\uparrow}^{} \rangle.
\end{equation}
Figure \ref{fig10} shows that this is indeed different from zero
within the phase 4. Interestingly the Hamiltonian does not
contain any attractive interaction between the conduction holes.
Rather, the singlet pairing between the localized spins as described
by the parameter $\Delta_{dd}$ is
`transferred' to the mobile conduction hole system by the
coherent Kondo pairing amplitude $\Delta_{cd}$. \\
Figure \ref{fig11} shows the gap and the $c$-like spectral weight
for several momenta along the $(1,0)$-direction and demonstrates
that the present theory qualitatively reproduces a well-known anomaly in 
cuprate superconductors: whereas the gap size near $(\pi,0)$
increases with decreasing $T_c$ - in contrast to what one expects
from BCS theory - the spectral weight decreases to zero\cite{Feng,Ding}.
At least the behaviour in Fig. \ref{fig11} is
easy to understand: the gap size around $(\pi,0)$ is determined
by $\Delta_{dd}$, the $c$-like spectral weight by $\Delta_{cd}$ which
governs the degree of mixing between $c$- and $d$-like bands.
Lower $T_c$ implies a lower value of $\Delta_{cd}$ at low $T$ and - since the 
two parameters compete with each other - a larger value of $\Delta_{dd}$.
\begin{figure}
\includegraphics[width=\columnwidth]{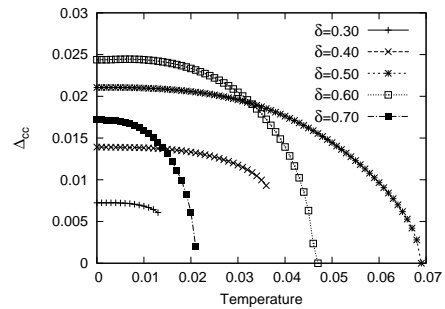}
\caption{\label{fig10}  
Pairing amplitude for conduction holes $\Delta_{cc}$ as a function
of temperature, $W/t=2.0$, $J/t=0.325$.}
\end{figure}
\begin{figure}
\includegraphics[width=\columnwidth]{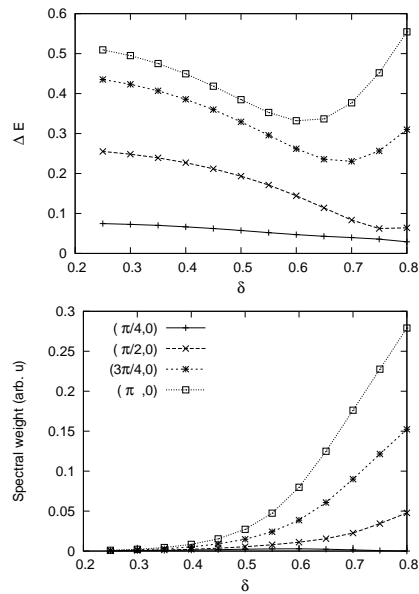}
\caption{\label{fig11}  
Energy gap (top) and spectral weight of the band below
$\mu$ (bottom) as a function of doping for some momenta
along $(1,0)$.
Parameter values are $W/t=2.0$, $J/t=0.325$, $T/t=0.05$.
}
\end{figure}
To conclude this section we briefly summarize the results
of the the mean-field theory: The competition between
Kondo-coupling and Heisenberg-exchange between localized spins
leads to a doping-driven phase transition between states
with different Fermi surface volume. The phase for high doping
is characterized by a complex order parameter which describes
coherent singlet-pairing between localized and conduction
holes and leads to a Fermi surface volume that includes
the localized spins. The low doping phase is characterized
by an order parameter which describes singlet correlations
between the localized spins and there is no coupling between conduction holes
and localized spins. The Fermi surface thus is a hole pocket
with a volume which does not include the localized spins.
Interestingly there is an intermediate phase where both
order parameters coexist. This phase has a gapped
large Fermi surface and shows superconducting correlations between
the conduction holes. These may be interpreted as the singlet correlations
between the localized spins being `transferred' to the conduction electrons
by the coherent Kondo singlet formation. This implies that the
superconducting transition has a very different character on the underdoped
and overdoped side of the superconducting dome: whereas
on the overdoped side the transition looks fairly conventional, with a gap
operning in the large Fermi surface, on the underdoped side the transition
corresponds to the sudden emergence of a gapped large Fermi surface from
the high-temperature phase with a hole pocket. 
A somewhat discomforting
feature of the transition in the underdoped range is the fact that it
is 1$^{st}$ order.
Lastly we mention that the criticial temperatures and dopings in the present
study are very different from the cuprates
but this can be hardly a surpise because we are studying
a different model (single band Kondo lattice rather than Kondo-Heisenberg
model) and mean-field theories can not be expcted to produce
quantitatively correct transition temperatures anyway.
\section{Strong coupling theory - small Fermi surface}
\begin{figure}
\includegraphics[width=\columnwidth]{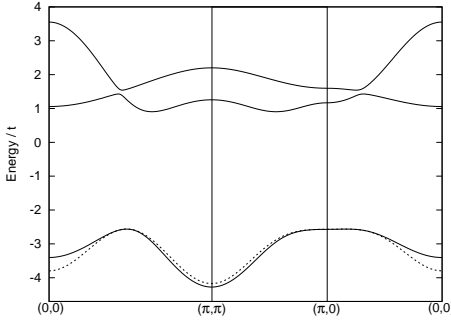}
\caption{\label{fig12}  
Dispersion of the quasiparticle bands for the phase with small
Fermi surface. Parameter values are $W/t=4$, $J/t=4$,
$\chi_{10}=-0.28$. The dashed line is the dispersion of the
lowest band as obtained by 2$^{nd}$ order perturbation theory.
}
\end{figure}

The mean-field theory in the preceding section may be 
expected to be good for
$t\gg W,J$ although even there relaxing the constraint on the occupation
of the $d$-orbital is problematic. The phsical regime of parameters, however,
is rather $W \gg t \gg J$. 
In fact a considerable deficiency of the mean-field treatment
lies the fact that in the hole pocket phase the
$d$- and $c$-holes are completely decoupled from one another, 
which clearly is unrealistic for large $W/t$.
In the following sections we therefore 
give a description of the two phases in a strong-coupling picture
which may be expected to hold best in the limit $W/t \gg 1$.
The discussion of the small-Fermi-surface phase in the present section
thereby closely follows the theory for the lightly doped Mott insulator
given in Ref. \cite{pockets}.  \\
The essence of the strong coupling theory is the approximate
diagonalization of the Hamiltonian in a suitably chosen truncated
Hilbert space. To construct this truncated
Hilbert space for the small-Fermi surface state
we start from  the case $\delta=0$ and consider
a state $|\Psi_0\rangle$ of the $d$-spin system which
has exactly one electron/site, is invariant under
point group operations, has momentum zero and is a spin singlet.
These are the quantum numbers of a vacuum state and indeed
$|\Psi_0\rangle$ will play the role of the vacuum state in our analysis.
The only property of $|\Psi_0\rangle$ 
which is relevant for the quasiparticle dispersion and total energy
is the static spin correlation function
\begin{equation}
\chi_{ij} = \langle \Psi_0 | {\bf S}_i \cdot {\bf S}_j |  \Psi_0 \rangle.
\end{equation}
We consider $\chi_{ij}$ as a given input parameter.
We assume it to be antiferromagnetic and of short range i.e.
\begin{equation}
\chi_{ij}= C_0 \;e^{i {\bf Q} \cdot({\bf R}_i - {\bf R}_j)}
 e^{-\frac{|{\bf R}_i - {\bf R}_j |}{\zeta}}
\label{chi1}
\end{equation}
where ${\bf Q}=(\pi,\pi)$. 
A more detailed discussion is given in Ref. \cite{pockets}.
It will be seen below that within the framework of the
present theory only the nearest neighbor
spin correlation $\chi_{10}$ has any relevance for the results.\\
Next, we define the following operators which add a conduction
hole to the system:
\begin{eqnarray}
\hat{a}_{i,\uparrow}^\dagger &=& \frac{1}{\sqrt{2}}(
\hat{c}_{i,\uparrow}^\dagger \hat{N}_{i\downarrow} - 
\hat{c}_{i,\downarrow}^\dagger S_i^+),
\nonumber \\
\hat{b}_{i,1,\uparrow}^\dagger &=& \frac{1}{\sqrt{2}}(
\hat{c}_{i,\uparrow}^\dagger \hat{N}_{i\downarrow} + 
\hat{c}_{i,\downarrow}^\dagger S_i^+),
\nonumber \\
\hat{b}_{i,2,\uparrow}^\dagger&=& \hat{c}_{i,\uparrow}^\dagger 
\hat{N}_{i\uparrow}.
\label{hubs}
\end{eqnarray}
Here capital/small letters are used for localized/conduction
holes, $\hat{c}_{i,\sigma}=c_{i,\sigma}(1-n_{\bar{\sigma}})$, 
and $\hat{N}_{i,\sigma}=N_{i,\sigma}(1-N_{\bar{\sigma}})$.
The operators (\ref{hubs}) add a conduction hole to the system,
thereby change the 
$z$-component of the total spin by $+1/2$ and
produce either a local singlet ($\hat{a}_{i,\uparrow}^\dagger$)
or one of the two components of a local triplet 
($\hat{b}_{i,1,\uparrow}^\dagger$ and $\hat{b}_{i,2,\uparrow}^\dagger$).
Analogous operators which change the
$z$-component of the total
spin by $-\frac{1}{2}$ are easily constructed.  \\
Using these operators we can now write down
the basis states of the truncated  Hilbert as
\begin{eqnarray}
2^{(N_\nu + N_\mu+N_\lambda)/2}&& \left(\prod_{\nu=1}^{N_\nu}\nonumber \hat{a}_{i_\nu,\sigma_\nu}^\dagger\;\right)\nonumber \\ 
&&\left(\prod_{\mu=1}^{N_\mu}  \hat{b}_{i_\mu,1,\sigma_\mu}^\dagger \;\right)
\left(\prod_{\lambda=1}^{N_\lambda}  \hat
{b}_{i_\lambda,2,\sigma_\lambda}^\dagger\;\right)
|  \Psi_0 \rangle.\nonumber \\ 
\label{basis_1}
\end{eqnarray}
Thereby it is understood that all sites $(i_\nu,j_\mu,k_\lambda)$
are pairwise different from each other, that means no two operators 
are allowed to act 
on the same site. The states (\ref{basis_1}) have singlets and triplets
at specified positions and we will treat these 
as spin-$\frac{1}{2}$
Fermions with a hard-core constraint, which is the key approximation of
the theory. Fermions are the only meaningful
description for these particles because operators of the type
(\ref{hubs}) 
which refer to different sites anticommute. Since 
$\langle \Psi_0 |\hat{a}_{i,\uparrow}	\hat{a}_{i,\uparrow}^\dagger |  
\Psi_0 \rangle
= \langle \Psi_0 |\hat{b}_{i,1,\uparrow} \hat{b}_{i,1,\uparrow}^\dagger |  
\Psi_0 \rangle
= \langle \Psi_0 |\hat{b}_{i,2,\uparrow} \hat{b}_{i,2,\uparrow}^\dagger |  
\Psi_0 \rangle
=\frac{1}{2}$
the states (\ref{basis_1}) are approximately normalized.
The issue of the normalization of the states has been discussed
in detail in Ref. \cite{pockets} where it was concluded that
this normalization
will be a good approximation in the
limit of short spin correlation length $\zeta$, 
that means the case of a `spin liquid' which is
the case of interest e.g. in the underdoped cuprates.
A more detailed discussion of this issue and others
is given in Appendix A.\\
As already stated we treat the singlets and triplets
as Fermionic quasiparticles i.e. 
the states (\ref{basis_1}) are represented by states
of Fermionic spin-$\frac{1}{2}$ quasiparticles
\begin{equation}
\left( \prod_{\nu=1}^{\hat{N}_\nu} a_{i_\nu,\sigma_\nu}^\dagger\;\right)
\left( \prod_{\mu=1}^{\hat{N}_\mu}  b_{i_\mu,1,\sigma_\mu}^\dagger \;\right)
\left( \prod_{\lambda=1}^{\hat{N}_\lambda}  
b_{i_\lambda,2,\sigma_\lambda}^\dagger\;\right)
|  \Psi_0 \rangle.
\label{basis_2}
\end{equation}
Operators in the quasiparticle Hilbert space 
then are defined by demanding that their matrix elements between
the states (\ref{basis_2}) are identical to those of the physical operators 
between the corresponding states (\ref{basis_1}). \\
We illustrate  this by setting up the quasiparticle Hamiltonian.
Straightforward calculation
gives the following matrix elements in the physical Hilbert
space:
\begin{eqnarray}
\langle \Psi_0 | \hat{a}_{j,\uparrow}  \;H_t\;\hat{a}_{i,\uparrow}^\dagger
|\Psi_0 \rangle 
&=& -t_{ij}(\frac{1}{8}+\frac{\chi_{ij}}{2})
\nonumber\\
\langle \Psi_0 | \hat{b}_{j,1,\uparrow}  \;H_t\;\hat{a}_{i,\uparrow}^\dagger
|\Psi_0 \rangle 
&=& -t_{ij}(\frac{1}{8}-\frac{\chi_{ij}}{6})
\nonumber\\
\langle \Psi_0 | \hat{b}_{j,2,\uparrow}  \;H_t\;\hat{a}_{i\uparrow}^\dagger
|\Psi_0 \rangle 
&=& -t_{ij}(\frac{1}{4\sqrt{2}}-\frac{\chi_{ij}}{3\sqrt{2}})
\nonumber\\
\langle \Psi_0 | \hat{b}_{j,1,\uparrow}  \;H_t\;\hat{b}_{i,1,\uparrow}^\dagger
|\Psi_0 \rangle 
&=& -t_{ij}(\frac{1}{8}+\frac{\chi_{ij}}{2})
\nonumber\\
\langle \Psi_0 | \hat{b}_{j,2,\uparrow}
\;H_t\;\hat{b}_{i,1,1\uparrow}^\dagger
|\Psi_0 \rangle \begin{large} 
\end{large}
&=& -t_{ij}(\frac{1}{4\sqrt{2}}+\frac{\chi_{ij}}{3\sqrt{2}})
\nonumber\\
\langle \Psi_0 | \hat{b}_{j,2,\uparrow}  \;H_t\;\hat{b}_{i,2,\uparrow}^\dagger
|\Psi_0 \rangle 
&=& -t_{ij}(\frac{1}{4}+\frac{\chi_{ij}}{3})
\label{raw}
\end{eqnarray}
They represent the decomposition of hopping into events of annihilation and 
creation of singlets and triplets.
The hopping term $H_t$ for the quasiaparticles thus takes the form
\begin{equation}
 H_t=\sum_{i,j} \sum_{\sigma} {\bf v}_{i,\sigma}^{\dagger} {\bf T}_{ij}  
{\bf v}_{j,\sigma},
\end{equation}
where
\begin{equation}
 {\bf v}_{i,\sigma}^{\dagger}=(a_{i,\sigma}^\dagger,b_{i,1,\sigma}^\dagger,
b_{i,2,\sigma}^\dagger),
\end{equation}
and 
\begin{equation}
{\bf T}_{ij}=t_{ij}\left(
\begin{array}{ccc}
\frac{1}{4}+\chi_{ij}&\frac{1}{4}-\frac{\chi_{ij}}{3} & \sqrt{2}
(\frac{1}{4}-\frac{\chi_{ij}}{3})\\
\frac{1}{4}-\frac{\chi_{ij}}{3}& \frac{1}{4}+\chi_{ij}&  \sqrt{2}
(\frac{1}{4}+\frac{\chi_{ij}}{3})\\
\sqrt{2}(\frac{1}{4}-\frac{\chi_{ij}}{3})& \sqrt{2}(\frac{1}{4}+
\frac{\chi_{ij}}{3}) & \frac{1}{2}+\frac{2\chi_{ij}}{3}
\end{array}
\right). \label{hopp}
\end{equation}
Thereby we have to keep in mind the hard-core constraint between the
quasiparticles. In addition it has been assumed that
`nearby' quasiparticles do not modify the matrix elements describing the
propagation of a given quasiparticle substantially. Again, this
will be justified in the limit of short spin correlation length $\zeta$.
\\
The key simplification is that in our restricted Hilbert space the
Kondo-coupling term $H_W$ takes the simple form
\begin{equation}
H_W=-\frac{3W}{4}\sum_{i,\sigma} a_{i\sigma}^\dagger a_{i\sigma}^{}
+\frac{W}{4}\sum_{i,\sigma} \sum_{n=1}^{2} b_{i,n,\sigma}^\dagger b_{i,n,\sigma}^{}.
\end{equation}
In the strong coupling limit $W/t\gg 1$
we thus expect to obtain something like one lower and two upper
`Hubbard bands' separated by an energy of order $W$.\\
It remains to represent $H_J$ in the new basis. Straightforward
computation gives the following matrix elements in the physical
Hilbert space:
\begin{eqnarray}
\langle \Psi_0 | \hat{a}_{i,\uparrow}  (\;{\bf S}_i \cdot {\bf S}_j) \;
\hat{a}_{i,\uparrow}^\dagger|\Psi_0 \rangle 
&=& 0
\nonumber\\
\langle \Psi_0 | \hat{b}_{i,1,\uparrow} \; ({\bf S}_i \cdot {\bf S}_j) \;
\hat{a}_{i,\uparrow}^\dagger|\Psi_0 \rangle 
&=& \frac{\chi_{ij}}{6}
\nonumber\\
\langle \Psi_0 | \hat{b}_{i,2,\uparrow} \; ({\bf S}_i \cdot {\bf S}_j) \;
\hat{a}_{i,\uparrow}^\dagger|\Psi_0 \rangle &=& \frac{\chi_{ij}}{3\sqrt{2}}
\nonumber\\
\langle \Psi_0 | \hat{b}_{i,1,\uparrow} \; ({\bf S}_i \cdot {\bf S}_j) \; 
\hat{b}_{j,1,\uparrow}^\dagger|\Psi_0 \rangle &=& 0
\nonumber\\
\langle \Psi_0 | \hat{b}_{i,2,\uparrow} \; ({\bf S}_i \cdot {\bf S}_j) \;
\hat{b}_{i,1,\uparrow}^\dagger|\Psi_0 \rangle &=& \frac{\chi_{ij}}{3\sqrt{2}}
\nonumber\\
\langle \Psi_0 | \hat{b}_{i,2,\uparrow} \; ({\bf S}_i \cdot {\bf S}_j) 
\;\hat{b}_{i,2,\uparrow}^\dagger|\Psi_0 \rangle &=& \frac{\chi_{ij}}{6}.
\label{rawhj}
\end{eqnarray}
so that the $d-d$ exchange for the quasiparticles takes the form
\begin{equation}
 H_J=z J \chi_{10} \sum_{i} \sum_{\sigma} {\bf v}_{i,\sigma}^{\dagger} 
K {\bf v}_{i,\sigma}+
\frac{zN J \chi_{10}}{2},
\label{exen}
\end{equation}
where $\chi_{10}$ denotes the nearest neighbor spin correlation function
and
\begin{equation}
{\bf K}=\left(
\begin{array}{ccc}
-1&\frac{1}{3} &\frac{\sqrt{2}}{3}\\
\frac{1}{3}& -1&  \frac{\sqrt{2}}{3}\\
\frac{\sqrt{2}}{3}& \frac{\sqrt{2}}{3} & -\frac{2}{3}
\end{array}
\right).
\end{equation}
The additional constant is the constribution from the
`spin background' whereby the correction due to the
$z$ broken bonds/quasiparticle is accounted for
in the first term.\\
The exchange term also has matrix elements of the type
\begin{equation} 
\langle \Psi_0 | \hat{b}_{i,1,\uparrow} \hat{b}_{j,1,\uparrow} \; ({\bf S}_i 
\cdot {\bf S}_j) \; \hat{a}_{i,\uparrow}^\dagger \hat{a}_{i,\uparrow}^\dagger
|\Psi_0 \rangle 
= \frac{1}{16}+\frac{\chi_{ij}}{12} \label{twobod}.
\end{equation}
Their contribution to the total energy will be $\propto \delta^2$
and we will neglect these.\\
Next we consider the hole count.
The number of localized holes always is $N$, the number of
sites in the system, whereas a self-evident
expression for the number of conduction holes is
\begin{equation}
N_c=\sum_{{\bf k},\sigma}\left(a_{{\bf k},\sigma}^\dagger a_{{\bf
    k},\sigma}^{}
+b_{{\bf k},1,\sigma}^\dagger b_{{\bf
    k},1,\sigma}^{}
+b_{{\bf k},2,\sigma}^\dagger b_{{\bf
    k},2,\sigma}^{}\right).
\end{equation}
This will give rise to a Fermi surface whose volume corresponds to
the doped holes but does not include the $d$-electrons.\\
The resulting quasiparticle band structure is shown
in Fig. \ref{fig12}. For a low density of quasiparticles it will 
be a reasonable
approximation to neglect the hard-core constraint,
because the probability that two particles occupy the same site
and thus violate the constraint is small. 
The Fermi surface then takes the form of a hole
pocket centered at ${\bf Q}=(\pi,\pi)$.
For a semi-quantitative discussion we may use 2$^{nd}$ order perturbation
theory for the dispersion $\epsilon_{1,{\bf k}}$
of the lowest quasiparticle band around ${\bf Q}$.
This gives
\begin{equation}
E_{\bf k}=-\frac{3W}{4} + \left(\frac{1}{4}+\chi_{10}\right) \epsilon_{\bf k}^{}
- \frac{3}{W}\left(\frac{1}{4}-\frac{\chi_{10}}{3}\right)^2\epsilon_{\bf k}^2
\label{smalldisp}
\end{equation}
The band minimum is at $(0,0)$ for $\chi_{10}>-\frac{1}{4}$
and at $(\pi,\pi)$ otherwise. The third term on the r.h.s. favours
a $\chi_{10}$ which is positive, the contribution
from the `spin background' favours a negative $\chi_{10}$.

Introducing ${\bf \kappa}={\bf k}-{\bf Q}$ or
${\bf \kappa}={\bf k}$, depending on whether the
minimum of the dispersion is at ${\bf k}=(\pi,\pi)$
or ${\bf k}=(0,0)$ we find
\begin{eqnarray}
\epsilon_{1,{\bf k}}&=&c_0 + c_1 {\bf \kappa}^2
\nonumber \\
c_0 &=& -\frac{3W}{4} \mp 4t\left(\frac{1}{4}+\chi_{10}\right) 
- \frac{48t^2}{W}\left(\frac{1}{4}-\frac{\chi_{10}}{3}\right)^2
\nonumber \\
c_1&=& \pm t\left(\frac{1}{4}+\chi_{10}\right)
+\frac{24t^2}{3W}
\left(\frac{1}{4}-\frac{\chi_{10}}{3}\right)^2
\label{small_sc}
\end{eqnarray}
and the ground state energy/site becomes
\begin{equation}
e_0 = c_0\delta + c_1 \pi \delta^2 + 2J\chi_{10}
\end{equation}
This can now be used to search for a minimum of
$e_0$ as a function of $\chi_{10}$. It turns out that
for most parameter values this expression favours
a $\chi_{10}$ which is as negative as possible.
We thus set $\chi_{10}=-0.33$, the value realized
in the ground state of the Heisenberg antiferromagnet.
This is obviously an absolute lower bound for the nearest neighbor
spin correlation function that can be realized by
any wave function.
\section{Strong coupling theory - large Fermi surface}
The strong coupling description of the phase with a large
Fermi surface has been given in Refs. \cite{kondo_1,kondo_2} - see also 
Ref.\cite{oestlund} for a different derivation -
and here we sketch the derivation only roughly.
We start again by defining the `vacuum state' for the theory
which now reads
\begin{equation}
|\Psi_0\rangle = 2^{-N/2}\;
\prod_i (c_{i,\uparrow}^\dagger d_{i,\downarrow}^\dagger
-c_{i,\downarrow}^\dagger d_{i,\uparrow}^\dagger) |0\rangle.
\label{vac}
\end{equation}
This is a product of local singlets and is the ground state
of the model for $t/W=J/W=0$ and a `hole doping' of $\delta=1$.
Acting with the hopping term $H_t$ onto (\ref{vac}) produces
charge fluctuations, i.e. states of the type
\begin{equation}
c_{j,\sigma}^\dagger c_{i,\sigma}^{}|\Psi_0\rangle.
\label{pair}
\end{equation}
In this state both cells, $i$ and $j$,
have a total spin of $1/2$ which is
carried by the unpaired $d$-hole-spin. We now identify
the quasiparticle-states of a single unit cell  $i$ as follows:
\begin{eqnarray}
|0\rangle &\rightarrow& \frac{1}{\sqrt{2}}(c_{i,\uparrow}^\dagger d_{i,\downarrow}^\dagger
-c_{i,\downarrow}^\dagger d_{i,\uparrow}^\dagger)|0 \nonumber \\
a_{i,\uparrow}^\dagger |0\rangle&\rightarrow& d_{i,\uparrow}^\dagger |0\rangle\nonumber \\
a_{i,\downarrow}^\dagger |0\rangle&\rightarrow& d_{i,\downarrow}^\dagger 
|0\rangle\nonumber \\
b_{i,\uparrow}^\dagger |0\rangle&\rightarrow& c_{i,\uparrow}^\dagger
c_{i,\downarrow}^\dagger d_{i,\uparrow}^\dagger |0\rangle\nonumber \\
b_{i,\downarrow}^\dagger |0\rangle&\rightarrow& c_{i,\uparrow}^\dagger
c_{i,\downarrow}^\dagger d_{i,\downarrow}^\dagger |0\rangle
\label{single_cell}
\end{eqnarray}
In this representation, the state (\ref{pair}) becomes
\begin{equation}
-\frac{1}{2}\; sign(\sigma)\;b_{j,\sigma}^\dagger\;
a_{i,\bar{\sigma}}^\dagger| 0\rangle.
\end{equation}
Just as (\ref{vac}), the state (\ref{pair})
is an eigenstate of $H_W$ with eigenvalue
$-(N-2)\frac{3W}{4}$. To keep track of this large
energy change we ascribe an energy of $3W/4$ to each of
the quasiparticles so that the representation of $H_W$ becomes
\begin{equation}
H_W=\frac{3W}{4}\;\sum_{i,\sigma}\;\left(a_{i,\sigma}^\dagger a_{i,\sigma}^{}
+b_{i,\sigma}^\dagger b_{i,\sigma}^{}\right).
\end{equation}
The particles are created/annihilated in pairs
and by subsequent application of the hopping term they also can propagate
individually. The procedure to be followed
then is entirely analogous as above:
we consider a quasiparticle Hilbert space whose basis is formed by
states of the type
\begin{equation}
\left(\prod_{\nu=1}^{N_a}\;a_{i_{\nu},\sigma_{\nu}}^\dagger\right)
\left(\prod_{\mu=1}^{N_b}\;b_{j_{\mu},\sigma_{\mu}}^\dagger\right)
|0\rangle.
\label{qp_large}
\end{equation}
As was the case for the small-Fermi-surface phase we assume
that the quasiparticles obey a hard-core constraint, i.e.
all sites $i_{\nu}$ and $j_{\mu}$ are pairwise different from each other.
The corresponding states in the Hilbert space of the physical
Kondo-lattice are
\begin{equation}
2^{-(N-N_a-N_b)/2}\;(-1)^{N_1}\;
\left(\prod_{\nu=1}^{N_a}\;c_{i_{\nu},\bar{\sigma}_{\nu}}^{}\right)
\left(\prod_{\mu=1}^{N_b}\;(-c_{i_{\mu},\sigma_{\mu}}^\dagger)\right)
|\Psi_0\rangle.
\label{qp_large_1}
\end{equation}
where 
\begin{eqnarray}
N_1=\sum_{\nu=1}^{N_a} \;\delta_{\sigma_{\nu},\downarrow}.
\end{eqnarray}
Again we construct operators for the quasiparticles by demanding that
the matrix elements of operators between the quasiparticle states
(\ref{qp_large}) are equal to those of the physical
operators between the corresponding Kondo-lattice states
(\ref{qp_large_1}). We thus obtain the Hamiltonian (see Ref. \cite{kondo_2}
for details)
\begin{eqnarray}
H &=& \frac{1}{2}\sum_{{\bf k},\sigma}
[\;(-\epsilon_{{\bf k}}+ \frac{3W}{2}) a_{{\bf k},\sigma}^\dagger
a_{{\bf k},\sigma}^{} +
(\epsilon_{{\bf k}}+\frac{3W}{2} )
b_{{\bf k},\sigma}^\dagger b_{{\bf k},\sigma}^{} \;]
\nonumber \\
&-& \frac{1}{2}\sum_{{\bf k},\sigma}
\; sign(\sigma)\;\epsilon_{{\bf k}} \;
(b_{{\bf k},\sigma}^\dagger a_{-{\bf k},\bar{\sigma}}^\dagger
 + H.c.).
\label{stcham}
\end{eqnarray}
where $\epsilon_{{\bf k}}=2t(\cos(k_x)+\cos(k_y))$ 
denotes the dispersion relation for the conduction holes.
If we assume that the density of quasiparticles is low -
which will hold true in the limit of large $W/t$ and close to
$\delta=1$ - it will again be a reasonable approximation to relax the hard core 
constraint. The Hamiltonian then can be diagonalized by the ansatz
\begin{eqnarray}
\gamma_{{\bf k},1,\sigma} &=& \;\;
u_{{\bf k}} b_{{\bf k},\sigma}^{}
+ v_{{\bf k}} sign(\sigma) a_{-{\bf k},\bar{\sigma}}^\dagger
\nonumber \\
\gamma_{{\bf k},2,\sigma} &=& -sign(\sigma) v_{{\bf k}} 
b_{{\bf k},\sigma}^{}
+ u_{{\bf k}} a_{-{\bf k},\bar{\sigma}}^\dagger
\label{ansatz}
\end{eqnarray}
and, introducing $\Delta$$=$$3W/2$, we
obtain the quasiparticle dispersion
\begin{eqnarray}
E_\pm({\bf k}) &=& \frac{1}{2}\;[\; \epsilon_{{\bf k}} \pm
\sqrt{ \epsilon_{{\bf k}}^2+ \Delta^2  }\;]
\nonumber \\
&\approx& \pm \frac{3W}{4} + \frac{\epsilon_{{\bf k}}}{2}
\pm\frac{\epsilon_{{\bf k}}^2}{4\Delta},
\label{scdisp}
\end{eqnarray}
where the second line holds in the limit $W/t \gg 1$.
The operator of total hole number is given by
\begin{eqnarray}
N_h&=&2N + \sum_{{\bf k},\sigma}\;\left(
b_{{\bf k},\sigma}^\dagger b_{{\bf k},\sigma}^{} -
a_{{\bf k},\sigma}^\dagger a_{{\bf k},\sigma}^{}\right)
\nonumber \\
&=&
\sum_{{\bf k},\sigma} \sum_{\mu=1}^2
\gamma_{{\bf k},\mu,\sigma}^\dagger \gamma_{{\bf k},\mu,\sigma}^{}.
\label{count_sc}
\end{eqnarray}
The first line follows from the fact that
the vacuum state (\ref{vac}) contributes $2N$ holes,
and that each hole-like/electron-like quasiparticle increases/decreases
the number of holes by one.
The second line shows that the lower of the two quasiparticle
bands is filled such that the Fermi surface has a volume
which corresponds to conduction holes {\em and} localized spins,
i.e. this state has a large Fermi surface.
The {\em apparent} contribution of the localized spins to the
Fermi surface can be understood as follows: at a conduction
hole density of $1$/unit cell the state resulting from the above
construction has a hole number of $2$/unit cell. At the same
time the state has an energy gap of order $3W/2$, the energy required to break
two singlets.
As far as the hole number and Fermi surface - or rather: absence
of a Fermi surface - are concerned, this state therefore is
completely equivalent to a band insulator, provided the localized spins 
contribute to the total hole number.
And since the quasiparticles introduced by changing $\delta$ are
spin-1/2 Fermions the Fermi surface at lower conduction hole
density is the same as that of a `doped band insulator'
and the localized spins {\em apparently} contribute to the Fermi surface
volume. This is therefeore simply the consequence of the fact that the
quasiparticles are spin-1/2 Fermions.\\
A notable feature of the above theory is, that it
actually incorporates the kind of broken gauge symmetry
which became apparent already in the mean-field treatment:
all single cell basis states of the phsical system
 - i.e. the states on the r.h.s. of (\ref{single_cell}) - are defined only 
up to a phase factor. The quasiparticle Hamiltonian 
(\ref{stcham}) then holds true only if
the phases of all states in a given unit cell are equal so that
no net phase enters the Hamiltonian.
However, one might choose e.g. the phase of the states
corresponding to $b_{i,\uparrow}^\dagger|0\rangle$ and 
$a_{i,\downarrow}^\dagger|0\rangle$ equal to unity and the phase of
the singlet state corrsponding to
$|0\rangle$ to be $exp(i\phi_j)$. In this case all matrix elements
in the real-space version of
(\ref{stcham}) would aquire extra phase factors. Such a phase-disordered
state might be adequate to describe the Kondo lattice at temperatures
lower than the Kondo temperature but higher than the coherence
temperature.
In the mean-field description we had a condensate
of singlets described by the constant order parameter $\Delta_{cd}$ - since the
constraint of localized $d$-holes cannot be taken into
account rigorously in mean-field theory this is probably the best
approximation to an array of phase-coherent Kondo singlets.\\
It remains to discuss the direct $d-d$ exchange $\propto J$.
In the vacuum state the $d-d$ exchange can promote two
singlets on neighboring sites into triplets. The matrix
element for this transition is $J/2$ so that
second order perturbation theory gives 
the correction to the energy/site
\begin{equation}
 -N \frac{z}{2}\frac{3J^2}{8W}.
\label{fluc}
\end{equation}
This is $\propto J^2/W$ and thus much smaller than the
direct $d-d$ exchange $\propto J$ in the phase with small Fermi surface
(see equation (\ref{exen})). This term will be even less important 
because this contribution occurs only if both cells
are unoccupied by quasiparticles.\\
In addition to this contribution of the vacuum there is also
a contribution of the quasiparticles to the exchange energy
because the spin of the quasiparticles
is carried by the $d$-electron.
For example we have
\begin{eqnarray}
S_i^+ &=& a_{i,\uparrow}^\dagger a_{j,\downarrow}^{}
+ b_{i,\uparrow}^\dagger b_{i,\downarrow}^{}
\end{eqnarray}
which again can be verified by comparing matrix elements
of the r.h.s. and l.h.s. between quasiparticle states and
physical states of the Kondo lattice.\\
If we relax the hardcore constraint a straightforward 
calculation then gives the contribution
to the ground state expectation value/site from the exchange between
quasiparticles of
\begin{eqnarray}
\frac{\langle H_J \rangle}{N} &=& -3J \; [
\left(\frac{1}{N}\sum_{\bf k}\gamma_{\bf k}v_{\bf k}^2f_{1,{\bf k}}\right)^2
+ \left(\frac{1}{N}\sum_{\bf k}\gamma_{\bf k}u_{\bf k}^2f_{1,{\bf k}}\right)^2
\nonumber \\
&& +2\left(\frac{1}{N}\sum_{\bf k}\gamma_{\bf k}u_{\bf k}v_{\bf k}
f_{1,{\bf k}}\right)^2\;]
\label{exch}
\end{eqnarray}
where $f_{1,{\bf k}}$ denotes the ground state occupation number
of the lower quasiparticle band (we have assumed that
the upper band is completely empty)
and $\gamma_{\bf k}=\frac{1}{2}(\cos(k_x)+\cos(k_y)$.
In the limit $W/t\gg 1$ we have $u_{\bf k}\rightarrow 1$,
$v_{\bf k}\rightarrow \frac{\sqrt{2}\epsilon_{\bf k}}{3W}$
so that we have approximately
\begin{equation}
\frac{\langle H_J \rangle}{N} = -3J
\left(\frac{1}{N}\sum_{\bf k}\gamma_{\bf k}f_{1,{\bf k}}\right)^2
\label{exch_1}
\end{equation}
Corrections to this will be of order $J/W$.\\
Whereas the small-Fermi-surface theory is valid for
electron densities close to $n=1$ electron/unit cell,
the large-Fermi-surface theory is valid for $n=2$ electron/unit cell.
In the next step we will compute the ground state energy as
a function of the electron density.
\section{Phase transition in strong coupling}
In the preceding two sections we have given a strong-coupling
description of the two phases with large and small Fermi surface.
Our main goal thereby was to elucidate how these phases can be
characterized beyond simple mean-field theory.
Clearly it would now be desirable
to compare the energies of the resulting ground states
and discuss a possible phase transition between the two. 
It should be noted from the very beginning that this will necessarily
involve some rather crude approximations because the quasiparticle Hamiltonians
for the two phases a formulated in terms of Fermions with a hard-core
constraint. The considerations in the following section thus will 
necessarily have a more qualitative character.\\
We first consider the dominant term in the Hamiltonian,
$H_W$ and collect only terms $\propto W$. 
In the phase with the small Fermi surface the number of quasiparticles 
is $N\delta$ and since each quasiparticle contributes
$-\frac{3W}{4}$ (see Eq. \ref{small_sc})
we have $\langle H_W \rangle = -\frac{3N\delta W}{4}+0(W^0)$.
The `vacuum' for the construction of the large Fermi surface state,
Eq. (\ref{vac}), contributes
an energy of $-N\frac{3W}{4}$. From the quasiparticle dispersion
(\ref{scdisp})
we obtain a contribution of $-N(1+\delta)\frac{3W}{4}+0(W^0)$ because the
total number of quasiparticles is
$N(1+\delta)$. And finally there is an additional
constant term of $2N\;\frac{3W}{4}$ which comes from inverting
the two $a$-Fermion operators in the first term on the
r.h.s. of Eq. (\ref{stcham}) so that we obtain
$\langle H_W \rangle = -\frac{3N\delta W}{4}$. 
To leading order in $W$ the two phases thus are
degenerate which is very different from the mean-field treatment. 
We thus consider the kinetic energy.
\begin{figure}
\includegraphics[width=\columnwidth]{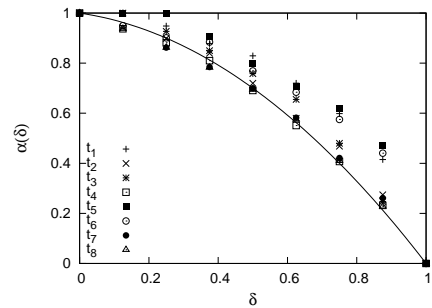}
\caption{\label{fig4}  
The function $\alpha(\delta)$ defined in (\ref{alphadef})
evaluated numerically
in a $4\times 4$ cluster for different forms of the kinetic
energy, i.e. for different values of the hopping integrals
$t$, $t'$ and $t''$. The values of the hopping integrals
for the individual systems are given in Table
\ref{tab1}. The line is the function (\ref{alfit}).
}
\end{figure}
In discussing the kinetic energy we encounter the problem that
the phase transition necessarily occurs for a doping range where
at least one of the two states with different Fermi surface
volume is far from `its' vacuum state (these correspond to
$\delta=0$ for the small Fermi surface state and
$\delta=1$ for the large Fermi surface state), so that
relaxing the respective hard-core constraint between the quasiparticles
will no longer be justified.
To obtain at least qualitative results, we proceed as follows:
we expect that for a finite density of quasiparticles
the hard-core constraint leads to a reduction of the total kinetic energy.
For a {\em single-band Hubbard model} we can study this reduction
by exact diagonalization of a small cluster.
More precisely, defining the ground state energy of
a single-band Hubbard model with Coulomb repulsion
$U$ and $n$ electrons by $E_0^{(n)}(U)$ we can
compute the function
\begin{equation}
\alpha(n/N) = \frac{E_0^{(n)}(U=\infty)}{E_0^{(n)}(U=0)}
\label{alphadef}
\end{equation}
where $N$ for the time being denotes the number of
sites in the cluster. This
gives the reduction of the kinetic energy due to
the hard-core constraint. By evaluating both energies
in the same $4\times 4$ cluster we may expect to cancel out
the shell-effects which inevitably dominate the kinetic energy
of a finite cluster. In the numerical computation of
$\alpha(n/N)$ we have moreover imposed the additional restriction
to use only states with total spin $S=0$.\\
We have performed this procedure for
different forms of the kinetic energy i.e. for different values
of the hopping integrals $t'$ and $t''$ to
$(1,1)$-like and $(2,0)$-like neighbors. 
\begin{table}[h,t]
\begin{center}
\begin{tabular}{|c|rrr|}
\hline
$H_t$ &  t  &   t'   &  t''  \\
\hline
 1 &   1.00 &   0.50 &   0.40 \\
 2 &   1.00 &  -0.50 &   0.40 \\
 3 &   1.00 &   0.50 &  -0.40 \\
 4 &   1.00 &  -0.50 &  -0.40 \\
 5 &   1.00 &   2.00 &   2.00 \\
 6 &   1.00 &  -2.00 &   2.00 \\
 7 &   1.00 &   2.00 &  -2.00 \\
 8 &   1.00 &  -2.00 &  -2.00 \\
\hline
\end{tabular}
\caption{The values of the hopping integrals for the calculation of
the renormalization of the kinetic energy in Figure \ref{fig4}.
The first column gives the number by which the dataset is labelled in 
Figure \ref{fig4}. }
\label{tab1}
\end{center}
\end{table}
Figure \ref{fig4} then shows the resulting $\alpha(\delta)$
for different combinations of $t'$ and $t''$.
The respective values of $t'$ and $t''$ for each
combination are given in Table \ref{tab1}.
As can be seen from Figure \ref{fig4}
$\alpha(\delta)$ is relatively independent on the precise form of the
kinetic energy, i.e. it seems to be an almost universal
function of the particle density. 
$\alpha(\delta)$ can be fitted quite well by a simple
quadratic function which depends on a single parameter
\begin{equation}
\alpha(\delta) = 1 + \lambda \delta - (1+\lambda) \delta^2
\label{alfit}
\end{equation}
where $\delta=n/N$. The fit gives the value $\lambda=-0.2$.\\
We now {\em assume} that the same reduction factor remains valid also
for the more complicated Hamiltonians produced by the above
strong-coupling theories. It should be noted that this cannot
be completely wrong because the values $\alpha(0)=1$ and
$\alpha(1)=0$ are known and the function may be expected to be
slowly varying near $\delta=0$ and rapidly varying near
$\delta=1$.\\
We therefore evaluate the kinetic energy for different
electron densities of the quasiparticles in the
states with large and small Fermi surface by computing the
kinetic energy in the absence of the hard-core constraint
and then correcting by the factor $\alpha(\delta)$.
Thereby for the large-Fermi-surface phase
$\delta$ is not the physical density of the conduction
holes, but the density of quasiparticles
\begin{equation}
\tilde{\delta}=\frac{1}{N}\; \sum_{{\bf k},\sigma}\left( 
a_{{\bf k},\sigma}^\dagger a_{{\bf k},\sigma}^{} +
b_{{\bf k},\sigma}^\dagger b_{{\bf k},\sigma}^{}\right).
\end{equation}
Let us stress that it is clear from the very beginning
that this procedure must lead to a level crossing for
a small value $\delta_c$ of the physical hole concentration:
in the absence of the correction factor $\alpha(\delta)$
the large-Fermi-surface-phase always has a lower
kinetic energy because the renormalization of the hopping integrals
is much weaker in this phase, compare Eqs. (\ref{stcham})
and (\ref{hopp}).
On the other hand, for $\delta\rightarrow 0$, the density
of the quasiparticles $\tilde{\delta}\approx 1-\delta$
for the large-Fermi-surface phase
approaches $1$, so that the correction factor $\alpha(\tilde{\delta})$ 
approaches zero, wheras the density of quasiparticles for the
small-Fermi-surface-phase is $\tilde{\delta}=\delta$ and
$\alpha(\tilde{\delta})$ is close to $1$. This gives
a level crossing 
even if only the kinetic energy is taken into account
and we moreover expect this level crossing to occur for small $\delta$.
\begin{figure}
\includegraphics[width=\columnwidth]{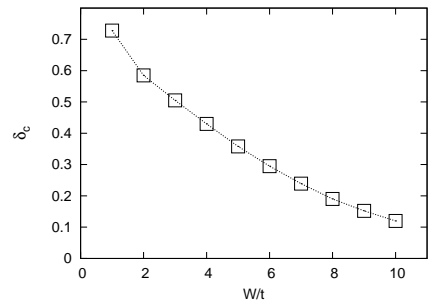}
\caption{\label{fig3}  
Hole concentration $\delta_c$ where the phase transition
from small to large Fermi surface occurs as a function
of $W/t$. The small Fermi surface is realized for
$\delta < \delta_c$.}
\end{figure}
Figure \ref{fig3} which shows $\delta_c$ obtained by numerical
calculation as a function of $W/t$ confirms this.
$\delta_c$ decreases with increasing $W/t$ which can be understood
as a consequence of the decrease of the bandwidth in the
small Fermi surface state with increasing $W/t$, compare (\ref{small_sc}).\\
Lastly, the $d-d$ exchange term $H_J$ should be considered but since we
obtain a phase transition already from the kinetic energy and since
the exchange energy is only a small correction for
physical parameters we restrict ourselves to a qualitative discussion.
For the small-Fermi surface phase we have the contribution
from the exchange energy of the localized spins
amongst each other, which was $2(1-2\delta)J\chi_{10}$.
The discussion in
section III suggested that $\chi_{10}$ is close to the
value for the Heisenberg antiferromagnet, which is a
lower bound for the nearest neighbor spin correlation function
of localized electrons.
The contribution from exchange between the quasiparticles,
see Eq. (\ref{twobod}) is
$\propto \delta^2$ and since the kinetic energy
favours a level crossing at small $\delta$ - at least for larger $W/t$ -
this contribution is not important.
For the large Fermi surface phase we first have a contribution
from the virtual pair creation of triplets, Eq. (\ref{fluc}).
Since this is of order $J^2/W$ and will be
suppressed because the density of quasiparticles is close to
$1$ this is not important either. Second, we have the
contribution (\ref{exch}) or (\ref{exch_1}) from the exchange between 
the quasiparticles themselves which is $\propto J$. Since there
the impact of the hard core constraint is hard to estimate
we cannot definitely say which phase is favoured by the Heisenberg
exchange - however, this is a small correction anyway.\\
All in all we thus expect a phase transition between the states
with small and large Fermi surface also in the strong coupling
description. With the approximations outlined above this
transition should occur for any value of $J/t$, which is different from
the mean-field theory where a minimum value of  $J/t$ was required.
Of course, quantitative agreement between mean-field and strong coupling 
theory may hardly be expected.
\section{Conclusion}
In summary, the doping induced transition
between phases with large and small Fermi surface in the 2D Kondo 
lattice model augmented by a Heisenberg exchange between the localized spins
has been studied by mean-field theory and in a strong coupling theory.
Mean-field theory produces a phase diagram which has a rough
similarity with that of cuprate superconductors:
for low doping and low temperature
the localized spins do not contribute to the Fermi surface
volume but form a decoupled spin liquid with pronounced singlet
pairing. The spin liquid corresponds to a bond-related
order parameter $\Delta_{dd}$ with d$_{x^2-y^2}$ symmetry
which describes singlet pairing between localized spins on
nearest neighbors. This phase likely corresponds to the pseudogap
phase in the cuprates. In mean-field theory the Fermi 
surface of the conduction holes is a pocket
around $\Gamma$ which is unphysical due
to the absence of any coupling to the localized spins. One might conjecture,
however, that the coupling to the antiferromagnetic fluctuations
of the spin liquid would create hole pockets near 
$(\pm\frac{\pi}{2},\pm\frac{\pi}{2})$ (see e.g. 
Refs. \cite{ederbecker,lauberciu})
as possibly observed in ARPES\cite{Damascelli,Meng}.
At higher doping and low temperature
the localized spins do contribute to the Fermi surface volume
and thus create a heavy Fermion-like state with a large
Fermi surface and an enhanced band mass. This corresponds to
the ovderdoped regime in the cuprates and is associated with
a complex on-site order parameter $\Delta_{cd}$, which describes
coherent local pairing between conduction electrons and localized 
spins.\\
At intermediate doping and low temperature
there is a phase where both order parameters coexist.
This phase appears to be a d$_{x^2-y^2}$ superconductor and has a large
Fermi surface with a d$_{x^2-y^2}$ gap. Superconductivity emerges because the
singlet pairing between the localized spins is transferred
to the mobile conduction holes by the coherent Kondo pairing.
All in all the phase diagram
thus shows a certain analogy with that of the cuprates.\\
In mean-field theory
the superconducting transition is very different in the underdoped
and overdoped regime: whereas in the overdoped regime we have
a conventional 2$^{nd}$ order transition with a d$_{x^2-y^2}$-like
gap opening on the large Fermi surface, the transition on the underdoped side
corresponds to the emergence of a gapped large Fermi surface
whereas the Fermi surface takes the form of hole pockets above the 
transition. This may actually be consistent with 
experiment\cite{ShenSawatzky}.
Mean-field theory moreover finds the transition to be 1$^{st}$
order on the underdoped side of the superconducting dome. 
This is on one hand not really expected for a 
superconducting transition but on the other hand
an example of a 1$^{st}$ order transition involving competing order 
parameters\cite{She}. If the transition really were first order
an interesting possibility emerges: namely if
the surface energy between the two degenerate phases were negative - as is the
case in type-II-superconductors - 
this would imply a tendency to form inhomogeneous states
in the underdoped region\cite{She}. In fact one peculiar feature of
underdoped cuprates is their 
`granularity'\cite{Pandavis,Langdavis,Pasupathy}.\\
One experimental feature which is not reproduced by the present theory
is the opening of a gap in an apparent large Fermi surface
at the pseudogap temperature as observed by Hashimoto 
{\em et al.}\cite{Hashimoto}. In the high-temperature phase the present
mean-field theory predicts a complete decoupling of localized and conduction 
holes. \\
To further elucidate the nature of the states with
different Fermi surface volume and the  existence of a phase
transition we have also performed a strong-coupling calculation.
In this theory, operators which create
the eigenstates of a single cell from a suitably chosen
`vacuum state' are treated as effective Fermions.
The advantage of this procedure is that the Kondo coupling, which should be
the largest energy scale in the problem, is treated essentially
exactly due to the `pre-diagonalization' of a single cell.
This theory then also yields a phase transition 
between states with different Fermi surface volume.
As opposed to the mean-field theory, the 
localized and conduction holes are approximately coupled to
a singlet in both phases, so that the expectation value of the 
Kondo exchange is the same for both phases.
In the phase with small Fermi surface realized at low doping
the Kondo singlets - or Zhang-Rice singlets in the case of cuprates -
form the quasiparticles so that the phase of a given Zhang-Rice singlet is
determined by its momentum ${\bf k}$ and it is these momenta which form
the small Fermi surface. In the large Fermi surface phase
the Kondo singlets have a uniform phase and the momenta which 
form the Fermi surface are carried by sites with 
spin $1/2$ and either $3$ or $1$ holes
(whereby the density of sites with $3$  holes is $\propto (W/t)^2$).\\
In both, the mean-field and the strong coupling theory, there occurs a
transition between the two states with different Fermi surface
for low carrier concentration.
Another complication which adds to the experimental complexity
and which has been addressed not at all in the present paper,
is a nematic ordering in the spin liquid plus hole pocket
phase. 
\acknowledgments
We thank  K. Grube for many instructive discussions.
R. E. most gratefully acknowledges the kind hospitality at the 
Center for  Frontier Science, Chiba University, Japan, where part of
this work was done.
\section{Appendix A}
In this Appendix we want to address various issues related to the
representation of the local singlets and triplets in terms of
spin-$1/2$ Fermion operators. We note first that
for example the states $\hat{a}_{i,\uparrow}^\dagger |  \Psi_0 \rangle$ and
$\hat{a}_{i,\downarrow}^\dagger |  \Psi_0 \rangle$ are 
orthogonal because their scalar product is proportional to
$\langle  \Psi_0|S_i^{+}  |  \Psi_0 \rangle = 0$.\\
Deviations from ideal Fermion behaviour appears in the
overlap of states with two Fermions. For example we have
\begin{eqnarray}
4\langle  \Psi_0 | \hat{a}_{i,\uparrow} \hat{a}_{j,\uparrow}^{}  
\hat{a}_{j,\uparrow}^\dagger  \hat{a}_{i,\uparrow}^\dagger |\Psi_0 \rangle
  &=& 1 + \frac{4}{3}\chi_{ij}\nonumber \\
4\langle  \Psi_0 | \hat{a}_{i,\uparrow} \hat{a}_{j,\downarrow}
\hat{a}_{j,\downarrow}^\dagger \hat{a}_{i,\uparrow}^\dagger  |\Psi_0 \rangle
  &=& 1 - \frac{4}{3}\chi_{ij}\nonumber \\
4\langle  \Psi_0 | \hat{a}_{i,\uparrow} \hat{b}_{j,1,\uparrow}^{}  
\hat{b}_{j,1,\uparrow}^\dagger  \hat{a}_{i,\uparrow}^\dagger |\Psi_0 \rangle
  &=& 1 + \frac{4}{3}\chi_{ij}\nonumber \\
4\langle  \Psi_0 | \hat{a}_{i,\uparrow} \hat{b}_{j,1,\downarrow}^{}  
\hat{b}_{j,1,\downarrow}^\dagger  \hat{a}_{i,\uparrow}^\dagger |\Psi_0 \rangle
  &=& 1 - \frac{4}{3}\chi_{ij}\nonumber \\
4\langle  \Psi_0 | \hat{a}_{i,\uparrow} \hat{b}_{j,2,\uparrow}^{}  
\hat{b}_{j,2,\uparrow}^\dagger  \hat{a}_{i,\uparrow}^\dagger |\Psi_0 \rangle
  &=& 1 - \frac{4}{3}\chi_{ij}\nonumber \\
4\langle  \Psi_0 | \hat{a}_{i,\uparrow} \hat{b}_{j,2,\downarrow}^{}  
\hat{b}_{j,2,\downarrow}^\dagger  \hat{a}_{i,\uparrow}^\dagger |\Psi_0 \rangle
  &=& 1 + \frac{4}{3}\chi_{ij}
\label{over}
\end{eqnarray}
The factors of $4$ on the left hand side thereby 
correspond to the prefactors in equation (\ref{basis_1}).
These relations would be consistent with those for ideal Fermi
operators if the spin correlation function $\chi_{ij}=0$.
If $\chi_{ij}$ is short ranged, the local singlets and triplets
thus behave like Fermion operators for `most distances'. We expect that 
the same will hold true for states with more than two Fermions provided
that they are pairwise more distant than the spin correlation length
$\zeta$. This will be a reasonable assumption in the limit of low density that
we are interested in. Furthermore, the neglected overlaps - 
being 'four particle overlaps' - would create an interaction between the 
quasiparticles rather than changing their dispersion.\\

\end{document}